\documentclass[pra,twocolumn,showpacs]{revtex4}
\usepackage{graphicx}
\usepackage{amsmath}
\usepackage{amssymb}
\usepackage{bm}

\begin{document}
\title{Bose-Einstein condensate as a quantum memory for a photonic polarization qubit}
\author{Stefan Riedl}
\altaffiliation{Present address: National Institute of Standards and Technology, 325 Broadway, Boulder, Colorado 80305, USA}
\author{Matthias Lettner}
\author{Christoph Vo}
\author{Simon Baur}
\author{Gerhard Rempe}
\author{Stephan D\"{u}rr}
\affiliation{Max-Planck-Institut f{\"u}r Quantenoptik, Hans-Kopfermann-Stra{\ss}e 1, 85748 Garching, Germany}


\hyphenation{re-corded}

\begin{abstract}
A scheme based on electromagnetically induced transparency is used to store light in a Bose-Einstein condensate. In this process, a photonic polarization qubit is stored in atomic Zeeman states. The performance of the storage process is characterized and optimized. The average process fidelity is $1.000 \pm 0.004$. For long storage times, temporal fluctuations of the magnetic field reduce this value, yielding a lifetime of the fidelity of $(1.1\pm0.2)$ ms. The write-read efficiency of the pulse energy can reach $0.53 \pm 0.05$.
\end{abstract}

\maketitle

\section{Introduction}

Optical quantum-memories \cite{lvovsky:09} based on electromagnetically induced transparency (EIT) \cite{fleischhauer:05} are a very active research area. Specifically, storage of classical light pulses \cite{liu:01, turukhin:01, phillips:01} and of single photons \cite{chaneliere:05, eisaman:05} was demonstrated. In addition, different polarizations of light were stored either using extended atomic level schemes \cite{matsukevich:06, tanji:09} or by converting the polarization into other degrees of freedom before storage \cite{choi:08, Cho:10, Zhang:11, England:1112.0900}. The vast majority of applications envisioned for quantum memories requires that quantum entanglement is first generated between two or more particles and that the quantum states of one or several of these particles are subsequently stored in a quantum memory. The crucial point is that this entanglement must survive the storage. Recently, this aspect was experimentally demonstrated in three independent experiments \cite{saglamyurek:11, clausen:11, lettner:11}.

Here we report in detail on the performance of the quantum memory used in one of these experiments \cite{lettner:11}. The experiment uses an $^{87}$Rb Bose-Einstein condensate (BEC) to realize a quantum memory for the polarization qubit of a single photon. A Raman scheme based on EIT is used to implement storage and retrieval of the photon. The atomic level-scheme is extended to allow for storage of the photonic polarization qubit in two atomic spin states. Quantum process tomography is used to determine the process fidelity which quantifies how well the polarization is maintained during storage. In addition, the decay of the process fidelity with increasing storage time is monitored.

All experiments reported here use classical light pulses instead of single photons, thus profiting from count rates which are much higher than in Ref.\ \cite{lettner:11}. These increased count rates yield a more precise value of the process fidelity. Ref.\ \cite{lettner:11} reports that storage and retrieval cause no discernible deterioration of the fidelity of the entangled state within an error bar of several percent. The present experiment still observes a process fidelity compatible with unity, but now with an error bar that is an order of magnitude smaller, thus demonstrating more clearly the capabilities of the BEC as a quantum memory.

In Ref.\ \cite{lettner:11}, a single $^{87}$Rb atom in an optical high-finesse cavity was used to generate a triggered single photon in such a way that the photon's polarization qubit is entangled with the spin state of the single atom. This unparalleled source is combined with the BEC that serves as a quantum memory for the single photon. The BEC is well suited for this purpose because the absence of thermal motion allows for long storage times, the large optical depth allows for high write-read efficiencies, and excellent internal-state preparation allows for high-fidelity storage of a qubit in atomic spin states. The experiment thus combines two different systems, each ideally suited for its purpose. The resulting hybrid character of the system poses an experimental challenge because the dipole traps that hold the single atom and the BEC in place have depths of several millikelvin and several microkelvin, respectively. Due to the resulting ac-Stark shifts, the single photons generated from the single atom are blue detuned by typically 70 MHz relative to the free-space atomic resonance, whereas the ac-Stark shifts experienced by the BEC are negligible on this scale. In the present paper, we therefore experimentally study the efficiency of light storage in the regime of 70 MHz single-photon detuning. The regime of much larger single-photon detunings has been studied theoretically \cite{gorshkov:07, nunn:07, mishina:08, sheremet:10} and experimentally \cite{reim:10, England:1112.0900} before, but those results are not immediately applicable to our system.

The paper is structured as follows: Section \ref{sec-implementation} describes the experimental implementation, Sec.\ \ref{sec-qubit} shows how well the polarization is maintained during storage, and Sec.\ \ref{sec-efficiency} studies the write-read efficiency. Appendix \ref{app} presents a simple model for coarsely estimating the write-read efficiency.

\section{Experimental implementation}
\label{sec-implementation}

\subsection{Electromagnetically induced transparency}
\label{sub-impl-EIT}

\begin{figure}[t!]
\includegraphics[width=0.75\columnwidth]{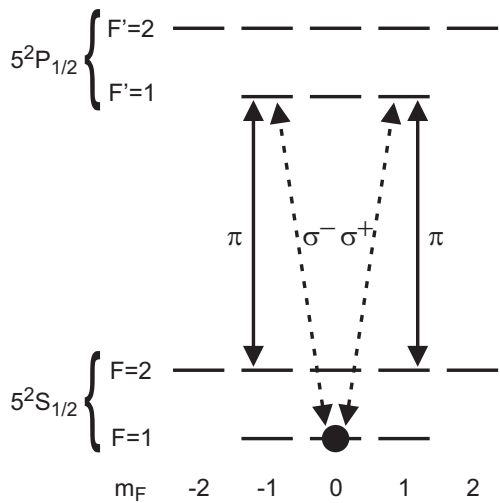}
\caption{
\label{fig-levels}
Atomic level scheme of the $D_1$ line in $^{87}$Rb. Probe light (dashed arrows) with an arbitrary superposition of the polarizations $\sigma^+$ and $\sigma^-$ couples the initial population ($\bullet$) in the hyperfine ground state $|F,m_F\rangle=|1,0\rangle$ to the $D_1$-line excited states $|F',m_F'\rangle=|1,\pm1\rangle$. $\pi$-polarized control light (solid arrows) couples these states to the hyperfine ground states $|F,m_F\rangle=|2,\pm1\rangle$. This makes it possible to store the probe-light polarization qubit in the qubit space spanned by the atomic states $|F,m_F\rangle=|2,\pm1\rangle$.
}
\end{figure}

EIT employs a control light beam to manipulate the propagation of a probe light beam inside a medium. Light storage in EIT-based schemes relies on the fact that the group velocity of the probe light $v_\mathrm{gr}$ can be reduced compared to the vacuum speed of light $c$ by many orders of magnitude \cite{hau:99} by choosing a small value for the control intensity. Upon entering the medium, the temporal duration of the probe pulse remains unchanged, whereas its spatial length is drastically reduced due to the small group velocity. A pulse which in vacuum is much longer than the medium can thus be fully compressed into the medium.

Once the pulse is fully inside the medium, one can ramp the control intensity to zero in an adiabatic fashion. In our experiment, we implement an approximately linear, temporal ramp of the control intensity which lasts 30 ns. This is sufficiently adiabatic according to Ref.\ \cite{fleischhauer:02}. This ramp reduces $v_\mathrm{gr}$ all the way to zero and the pulse is stopped inside the medium. The pulse is stored for a time $t_\mathrm{store}$ which can be chosen freely. After this, we ramp the control intensity back on and the pulse resumes its propagation \cite{liu:01, turukhin:01, phillips:01, fleischhauer:00}. During the storage time $t_\mathrm{store}$, the pulse exists in the medium in the form of an atomic spin wave. If the light pulse is compressed such that it fits inside the medium, then the spin wave stores the longitudinal and transverse shape of the light pulse.

\subsection{Atomic level scheme}
\label{sub-impl-levels}

Figure \ref{fig-levels} shows the atomic level scheme used in our experiment. Control and probe light for EIT are both resonant with the atomic $D_1$ line of $^{87}$Rb at a wavelength of $\lambda=795$ nm. The atoms are initially prepared in the hyperfine ground state $|F,m_F\rangle=|1,0\rangle$. The $\sigma^\pm$ polarized components of the probe light couple this population to the excited hyperfine states $|F',m_F'\rangle=|1,\pm 1\rangle$. The $\pi$-polarized control light transfers this population to the ground hyperfine states $|F,m_F\rangle=|2,\pm1\rangle$.

\subsection{Optical beam path}
\label{sub-impl-beam-path}

\begin{figure}[t!]
\includegraphics[width=0.95\columnwidth]{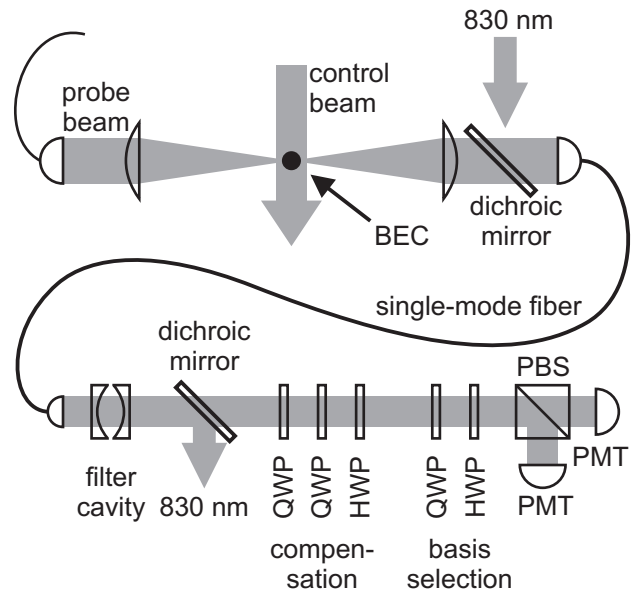}
\caption{
\label{fig-beam-path}
Simplified scheme of the optical beam path. A detailed description is given in the text.
}
\end{figure}

Figure \ref{fig-beam-path} shows a simplified scheme of the optical beam path. An $^{87}$Rb BEC serves as a quantum memory. The BEC is illuminated by $\pi$-polarized control light propagating along the $y$ axis with a waist ($1/e^2$ radius of intensity) of $\sim 100$ $\mu$m which is much larger than the Thomas-Fermi radii of the BEC so that the control light intensity can be approximated as constant across the BEC. In addition, the BEC is illuminated by probe light propagating along the $z$ axis focused to a waist of 8 $\mu$m. This is comparable to the Thomas-Fermi radii so that the probe beam samples some fraction of the transverse inhomogeneity of the BEC. To obtain a well-defined transverse mode for the probe light before impinging onto the BEC, the light is sent through a single-mode fiber. The polarization of the probe light can be $\sigma^+$, $\sigma^-$, or any superposition thereof.

After storing and retrieving the probe light, we need to measure its polarization. To this end, the beam path ends with a polarizing beam splitter (PBS) cube and two identical detectors, one in each output port of the PBS. A quarter-wave plate (QWP) followed by a half-wave plate (HWP), both placed right in front of the PBS allow for the selection of an arbitrary polarization basis. As detectors, we use photomultiplier tubes (PMTs) in this paper, instead of the avalanche photodiodes that we used in Ref.\ \cite{lettner:11}.

\subsection{Stray light filtering}
\label{sub-impl-filtering}

Stray light is an issue in our setup. Much of it is eliminated using mechanical shielding and temporal gating of the detector signals. The remaining stray light is dominated by control light off-resonantly scattered from the BEC during the retrieval of the probe pulse.

This stray light level would be unproblematic for the measurements presented here, but the experiments reported in Ref.\ \cite{lettner:11} required a substantial suppression. In the beam path from the BEC to the detectors, our setup therefore includes a single-mode optical fiber for transverse mode filtering and a filter cavity for spectral filtering.

The spatial filtering with the single-mode fiber makes use of the fact that storage and retrieval have little effect on the transverse mode of the probe light, whereas the control light is off-resonantly scattered from all positions in the BEC and into all directions. The fiber reduces the stray light power that reaches the detector by a factor of 0.068. In the absence of the BEC, the fiber reduces the probe light power by a factor of 0.66. Hence, the fiber increases the signal-to-background ratio by one order of magnitude. Storage and retrieval in the BEC compromise the transverse mode of the probe beam slightly. This causes an additional reduction of the probe light power by a factor of 0.88 at the single-photon resonance and by a factor of 0.80 for a single-photon detuning of $\Delta_c=2\pi\times 70$ MHz. In addition, the fiber suppresses the excitation of higher transverse modes in the subsequent filter cavity.

The filter cavity is a near-planar, plane-concave Fabry-Perot resonator with a finesse of 180 and a free spectral range of 40 GHz. The cavity has a transmission of 0.8 at the resonance, which is tuned to the probe light frequency. We expect that the scattered EIT control light is either elastically scattered or Raman scattered, thereby transferring an atom from $F=1$ to $F=2$. Hence, we expect the scattered light to be red detuned from the probe light by either 6.8 GHz or 13.6 GHz. Transmission through the cavity suppresses the light power for each of these frequencies by a factor of $2\times10^{-4}$. The cavity length is stabilized against long-term drift with a piezo actuator using a Pound-Drever-Hall technique with light at a wavelength of 830 nm. This light is overlapped with the probe light on a dichroic mirror in front of the mode-filtering fiber. The 830 nm light transmitted through the filter cavity is separated from the probe light using another dichroic mirror behind the filter cavity. A small fraction of the 830 nm light keeps propagating towards the detectors. This light is removed with dielectric interference filters which are not shown in Fig.\ \ref{fig-beam-path}. In addition, the 830 nm light source is turned off during any time intervals where relevant detector signals are expected.

The optical fiber for stray-light filtering is not polarization maintaining because it must work equally well for all possible input polarizations. Hence, the polarization after transmission through the fiber is related to the input polarization by a unitary transformation. As long as the fiber is not moved mechanically or exposed to temperature changes, this transformation is temporally stable. The same applies to the polarization transformations caused by the filter cavity, by the dichroic mirrors, and by other mirrors in the beam path which are not shown in Fig.\ \ref{fig-beam-path}. We compensate the resulting overall transformation using two QWPs followed by a HWP --- a combination which can generate any unitary transformation.

\subsection{Preparation of the BEC}
\label{sub-impl-BEC}

\begin{figure*}[t!]
\includegraphics[width=0.9\textwidth]{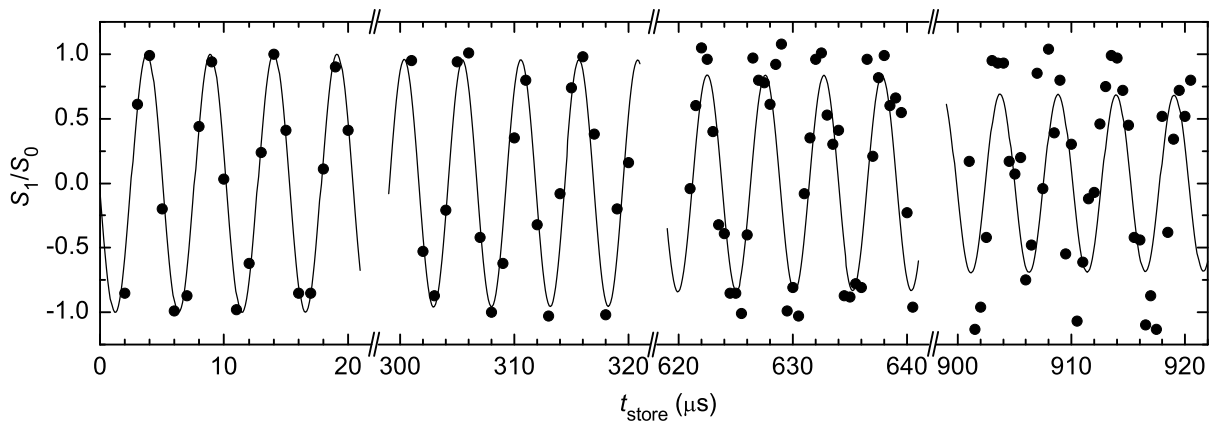}
\caption{
\label{fig-Faraday}
Faraday rotation. The normalized Stokes parameter $S_1/S_0$ oscillates as a function of storage time due to an applied magnetic hold field. The line is a fit of Eq.\ \eqref{S-1}. The best-fit value for the $e^{-1/2}$ damping time is $\sigma_\alpha=(1.1\pm0.2)$ ms. Note the breaks on the horizontal axis.
}
\end{figure*}

We produce an almost pure BEC in the hyperfine state $|F,m_F\rangle=|1,-1\rangle$, using radio-frequency (rf) induced evaporative cooling in a magnetic trap, as described in Ref.\ \cite{marte:02}. The gas is transferred into a crossed-beam optical dipole trap operated at a wavelength of 1064 nm. The measured trap frequencies are $(\omega_x,\omega_y,\omega_z)/2\pi=(70,20,20)$ Hz with gravity pointing along the $x$ axis. A magnetic hold field of $\sim 1$ G applied along the $z$ axis preserves the spin orientation of the atoms.

We use two consecutive microwave pulses, each with a pulse area of $\pi$, to transfer the population into the internal state needed for our EIT level scheme. Starting from state $|1,-1\rangle$, the first pulse transfers the population into state $|2,0\rangle$. Subsequently, the second pulse transfers the population to state $|1,0\rangle$. The total process transfers $\sim 90$ \% of the atoms into state $|1,0\rangle$. Atoms left in the $F=2$ hyperfine states are then removed with blast light. This is followed by temporary application of a strong magnetic field gradient which removes atoms with $m_F\neq0$ from the shallow optical dipole trap. After this procedure, the total atom number in undesired internal states lies below the detection limit of our setup which we estimate to be $\sim 200$ atoms. For the rest of the experiment, the magnetic hold field is reduced to typically 0.1 G. At this point, the BEC typically contains $N=1.2\times 10^6$ atoms. The corresponding Thomas-Fermi radii are $(R_x,R_y,R_z)=(7,25,25)$ $\mu$m.

\section{A quantum memory for the polarization qubit}
\label{sec-qubit}

We now study how well the polarization of the probe light is maintained during storage. The probe beam propagates along the $z$ axis. Hence, an arbitrary incoming polarization state can be expanded as $c_+|\sigma^+\rangle + c_- |\sigma^-\rangle$ with coefficients $c_+$ and $c_-$. With the atomic level scheme shown in Fig.\ \ref{fig-levels}, this state is mapped onto the atomic state $c_+|2,+1\rangle + c_- |2,-1\rangle$. The retrieval process maps the atomic state back to the original polarization state. In our experiment, both mapping processes work extremely well, but magnetic field noise causes a deterioration of the state for long storage times.

We apply a magnetic hold field of $B_z\sim0.1$ G and orient it along the $z$ axis. This suppresses undesired transitions between different Zeeman states caused by components of the magnetic-field noise perpendicular to the $z$ axis. In Sec.\ \ref{sub-qubit-Faraday} we discuss the Faraday rotation caused by this hold field. In Sec.\ \ref{sub-qubit-tomography} we use quantum state tomography to characterize the deterioration of the polarization for long storage times and show how techniques that reduce the magnetic-field noise improve the performance of the system.

\subsection{Faraday rotation}
\label{sub-qubit-Faraday}

The polarization of the probe light can be characterized using the Stokes parameters \cite{hecht:01, brosseau:98}
\begin{subequations}
\label{Stokes}
\begin{eqnarray}
S_0 = I_H+I_V
, &&\qquad
S_1 = I_H-I_V
, \\
S_2 = I_D-I_A
, &&\qquad
S_3 = I_R-I_L
,\end{eqnarray}
\end{subequations}
where $I_H$, $I_V$, $\hdots$ denote the intensity detected after a polarizer that transmits only one polarization, namely horizontal $H$, vertical $V$, diagonal ($+45^\circ$) $D$, anti-diagonal ($-45^\circ$) $A$, right circular $R$, or left circular $L$. Here, $R$ and $L$ correspond to $\sigma^+$ and $\sigma^-$. The Stokes parameters can be regarded as the components of a four-dimensional Stokes vector. $S_0$ describes the total intensity, whereas the three-dimensional vector
\begin{eqnarray}
\bm u = \frac1{S_0} \left(\begin{array}{c}
S_1 \\ S_2 \\ S_3
\end{array}\right)
\end{eqnarray}
describes the polarization of the light and is well suited for graphical visualization, in close analogy to the Bloch vector. We call $\bm u$ the Poincar\'{e} vector. Its unit sphere is called Poincar\'{e} sphere.

The magnetic hold field along the $z$ axis gives rise to a Faraday rotation of the Poincar\'{e} vector around its $z$ axis at an angular frequency
\begin{eqnarray}
\omega_F = \frac{\mu_B g_F \Delta m_F}\hbar B_z
,\end{eqnarray}
where $\mu_B=2\pi\hbar \times 1.40$ MHz/G is the Bohr magneton and $g_F$ is the Land\'{e} factor. For the levels used in our experiment $g_F=1/2$ and $\Delta m_F=2$. The total rotation angle of the Poincar\'{e} vector
\begin{eqnarray}
\label{phi}
\phi = \omega_F (t_\mathrm{store} + \tau_d(L))
\end{eqnarray}
has one contribution $t_\mathrm{store}$ from the storage time with the control light off and another contribution $\tau_d(L)$ which is the delay of the probe pulse caused by the propagation through the medium of length $L$ with the control light on. Note that the rotation of the polarization vector of the electric field in real space is a factor of 2 slower than the rotation of the Poincar\'{e} vector.

Figure \ref{fig-Faraday} shows experimental data of this Faraday rotation for a linear input polarization. The line shows a fit of a sinusoid with Gaussian damping
\begin{eqnarray}
\label{S-1}
\frac{S_1}{S_0} = e^{-t_\mathrm{store}^2/2\sigma_\alpha^2} \cos (\phi-\phi_0)
,\end{eqnarray}
where $\sigma_\alpha$ is the $e^{-1/2}$ damping time, where $\phi$ is given by Eq.\ \eqref{phi}, and where $\phi_0$ represents the input polarization. The best-fit values are $\omega_F=2\pi \times 0.20$ MHz, yielding $B_z=0.14$ G, and $\sigma_\alpha=(1.1\pm0.2)$ ms.

A careful inspection of the experimental data points for long times in Fig.\ \ref{fig-Faraday} leads us to an interesting observation, revealing the physical origin of the damping. Unlike the best-fit curve, the data points, which represent a single experimental shot each, do not show a decrease of the peak-to-peak amplitude. Instead, they are noisy insofar as they do not form a smooth sinusoid. We attribute this to irreproducible, temporal fluctuations of $B_z$, which effectively scatter the data points along the horizontal axis. Only if we were to average several experimental shots to represent their mean value, would we observe a reduction of the peak-to-peak values in the experimental data.

A fit of $S_1/S_0=A\cos(\phi-\phi_0)$ to the data in Fig.\ \ref{fig-Faraday} with $t_\mathrm{store}\leq20$ $\mu$s yields a best-fit value of $A=1.02\pm0.04$. The fact that $A$ is consistent with 1 shows that temporal fluctuations of $B_z$ on a time scale of 20 $\mu$s or faster have no discernible effect. Temporal fluctuations of $B_z$ on slower time scales manifest themselves in our experiment only as shot-to-shot fluctuations of $\phi=(\mu_B g_F \Delta m_F/\hbar) \linebreak[1] \int_0^{t_\mathrm{store}+\tau_d(L)} dt \linebreak[1] B_z(t)$. In our experiment, each shot requires the preparation of a new BEC, which takes 20 s. This suggests that shot-to-shot fluctuations of $B_z$ probably yield an important contribution to the shot-to-shot fluctuations of $\phi$.

\subsection{Quantum process tomography}
\label{sub-qubit-tomography}

We now turn to a complete characterization of the effect which the storage and retrieval process has on the polarization. As the process does not conserve the total intensity, a full description of the process must use the four-dimensional Stokes vector, not just the three-dimensional Poincar\'{e} vector. We consider the regime of small probe intensity. Hence, the dependence of the Stokes parameters $S_i^\mathrm{out}$ of the retrieved probe pulse on the Stokes parameters $S_i^\mathrm{in}$ of the incoming probe pulse can be approximated as linear
\begin{eqnarray}
S_i^\mathrm{out} = \sum_j M_{ij} S_j^\mathrm{in}
.\end{eqnarray}
$M$ is called M\"{u}ller matrix \cite{hecht:01, brosseau:98}.

As shown in Fig.\ \ref{fig-beam-path}, we use a PBS with detectors behind both output ports. According to Eq.\ \eqref{Stokes}, a full characterization of the Stokes vector requires such measurements for 3 different settings of the wave plates in front of the PBS, which select the measurement basis. This set of 3 measurements fully characterizes the quantum state of the polarization and it can be regarded as quantum \emph{state} tomography \cite{nielsen:00}.

To determine $M$ experimentally, we use a set of 4 linearly independent input Stokes vectors (e.g., $H$, $D$, $R$, and $L$) and perform quantum state tomography of the output state generated for each input state. This set of 12 measurements allows for a complete determination of $M$ and it can be regarded as quantum \emph{process} tomography \cite{nielsen:00}.

Performing such quantum process tomography, we find that in our experiment the M\"{u}ller matrix is always well approximated by
\begin{eqnarray}
\label{M}
M = \eta \left(\begin{array}{cccc}
1 & 0 & 0 & 0 \\
0 & \alpha \cos \phi & -\alpha \sin \phi & 0 \\
0 & \alpha \sin \phi & \alpha \cos \phi & 0 \\
0 & 0 & 0 & 1 \\
\end{array}\right)
,\end{eqnarray}
where $\phi$ is the angle resulting from the Faraday rotation, $\alpha$ is a damping factor, and $\eta$ is the write-read efficiency which is experimentally found to be independent of the input polarization.

Above, we concluded from Fig.\ \ref{fig-Faraday} that the polarization at short storage times is essentially pure, whereas at long storage times shot-to-shot fluctuation of $\phi$ must be taken into account. Hence, each individual shot can be described by some realization of $M$ as in Eq.\ \eqref{M} with $\alpha=1$ and with some value of $\phi$ which exhibits shot-to-shot fluctuations. We assume that the values of $\phi$ have a Gaussian distribution with root-mean-square (rms) width $\sigma_\phi$. Averaging over many shots yields Eq.\ \eqref{M} with $\alpha=\exp(-\sigma_\phi^2/2)$.

To develop a simple model for the dependence of $\alpha$ on $t_\mathrm{store}$, we assume that only shot-to-shot fluctuations of $B_z$ contribute to the fluctuations of $\phi$, i.e.\ we approximate $B_z$ as constant during each individual shot. With this approximation, the values of $B_z$ will have a Gaussian distribution with rms width $\sigma_B$ and with $\sigma_\phi=(\mu_B g_F \Delta m_F/\hbar)\sigma_B t_\mathrm{store}$, where we neglected $\tau_d(L)\ll t_\mathrm{store}$. This yields
\begin{eqnarray}
\label{alpha-def}
\alpha = \exp(-t_\mathrm{store}^2/2\sigma_\alpha^2)
\end{eqnarray}
with
\begin{eqnarray}
\label{sigma-alpha}
\frac1{\sigma_\alpha} = \sigma_B \frac{\partial \omega_F}{\partial B_z}
.\end{eqnarray}
For a linearly polarized input state, Eqs.\ \eqref{M} and \eqref{alpha-def} reproduce Eq.\ \eqref{S-1}. The Faraday rotation is a unitary evolution. For any given $t_\mathrm{store}$, its effect can be compensated, e.g., using wave plates and it is therefore not much of a concern. The non-unitary damping $\alpha$, however, irreversibly deteriorates the performance of the memory.

\begin{figure}[t!]
\includegraphics[width=0.9\columnwidth]{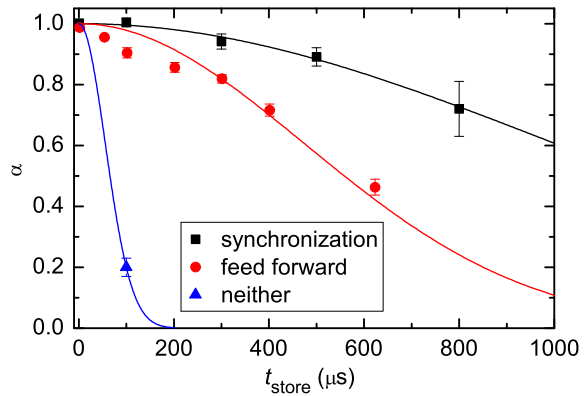}
\caption{
\label{fig-alpha}
(Color online) The damping factor $\alpha$ of Eq.\ \eqref{M} as a function of storage time. The lowest damping is obtained when synchronizing the start of the EIT write-read cycle with the 50 Hz ac line voltage ($\blacksquare$). Alternatively, we can use an open-loop feed-forward compensation ($\bullet$). Without any compensation, the damping is much stronger ($\blacktriangle$). The lines show Gaussian fits according to Eq.\ \eqref{alpha-def}.
}
\end{figure}

Experimental results for the time dependence of the damping parameter $\alpha$, as determined by quantum process tomography, are shown in Fig.\ \ref{fig-alpha}. Data taken without any reduction of magnetic-field noise ($\blacktriangle$) yield a value of $\sigma_\alpha = 0.06$ ms for the $e^{-1/2}$ damping time, corresponding to $\sigma_B=2$ mG according to Eq.\ \eqref{sigma-alpha}. In our experiment, the majority of this magnetic-field noise is periodic and in phase with the 50 Hz ac line voltage. We can suppress this noise drastically by synchronizing the start of the EIT write-read cycle with the ac line voltage. The corresponding data ($\blacksquare$) in Fig.\ \ref{fig-alpha} yield a best-fit value of $\sigma_\alpha = (1.0\pm0.1)$ ms, corresponding to $\sigma_B=0.1$ mG. Evidently, the synchronization improves $\sigma_\alpha$ by a factor of $\sim 20$. The data in Fig.\ \ref{fig-Faraday} were also taken with this synchronization and essentially reproduce the improved value of $\sigma_\alpha$.

Our experiments described in Ref.\ \cite{lettner:11} required a repetition of EIT write-read cycles at a rate of 10 kHz for a total time span of several seconds. Hence, write-read cycles had to occur at essentially all possible phases of the 50 Hz ac line voltage. To reduce the noise in these measurements, we first determined the values of $B_z(t)$ for one 50 Hz period in a series of calibration measurements. We then ran a current through a coil to compensate the recorded noise with an open-loop feed-forward circuit. The corresponding data ($\bullet$) in Fig.\ \ref{fig-alpha} yield a best-fit value of $\sigma_\alpha = (0.49\pm0.04)$ ms, corresponding to $\sigma_B=0.2$ mG. This compensation was good enough not to be the limiting factor in the overall experiment of Ref.\ \cite{lettner:11}, where we observed the same lifetime but with an error bar that was a factor of four larger.

The full information from the quantum process tomography is contained in $M$. To compare the overall performance of different quantum memories, one often uses the average process fidelity as a figure of merit. In terms of quantum states, the fidelity can be written as $F=\mathrm{Tr} (\rho_\mathrm{in}\rho_\mathrm{out})$, where Tr denotes the trace, $\rho$ denotes the density matrix, and we assumed that $\rho_\mathrm{in}$ represents a pure state. For polarization states, this can be rewritten as $F=(1+\bm u_\mathrm{in}\cdot \bm u_\mathrm{out})/2$. When averaging this quantity over all possible pure input states, i.e. over the surface of the Poincar\'{e} sphere, we obtain the average process fidelity $\langle F\rangle$ \cite{lvovsky:09, bowdrey:02}. After compensation of the Faraday rotation, i.e.\ for $\phi=0$, Eq.\ \eqref{M} yields
\begin{eqnarray}
\langle F\rangle= \frac13(2+\alpha)
.\end{eqnarray}
The synchronized data in Fig.\ \ref{fig-alpha} yield $\langle F \rangle = 1.000 \pm 0.004$ at $t_\mathrm{store}= 1$ $\mu$s and $\langle F \rangle = 0.90\pm0.02$ at $t_\mathrm{store}= 800$ $\mu$s. The value at $t_\mathrm{store}= 1$ $\mu$s shows that the state mapping between photonic and atomic qubit states works extremely well.

\section{Write-read efficiency}
\label{sec-efficiency}

Now, we turn to another important figure of merit for light storage, namely the efficiency of a complete write-read cycle
\begin{eqnarray}
\eta = \frac{E_\mathrm{retr}}{E_\mathrm{in}}
,\end{eqnarray}
defined as the energy of the retrieved probe pulse $E_\mathrm{retr}$ divided by the energy of the incoming probe pulse $E_\mathrm{in}$. For our memory, $\eta$ is independent of the polarization of the probe field, as seen in Eq.\ \eqref{M}. Hence, for understanding $\eta$ it suffices to consider the case where the probe polarization is fixed to $\sigma^+$. This simplifies the relevant atomic level scheme to a $\Lambda$-type three-level system. Unlike previous experiments by other groups, our work has a focus on the regime of 70 MHz detuning from the single-photon resonance.

We experimentally study the dependence of $\eta$ on the intensity of the control laser in Sec.\ \ref{sub-eta-omega-C}. In Sec.\ \ref{sub-eta-time}, we study the decay of the efficiency for long storage times.

\subsection{Dependence on the control intensity}
\label{sub-eta-omega-C}

\begin{figure}[t!]
\includegraphics[width=0.9\columnwidth]{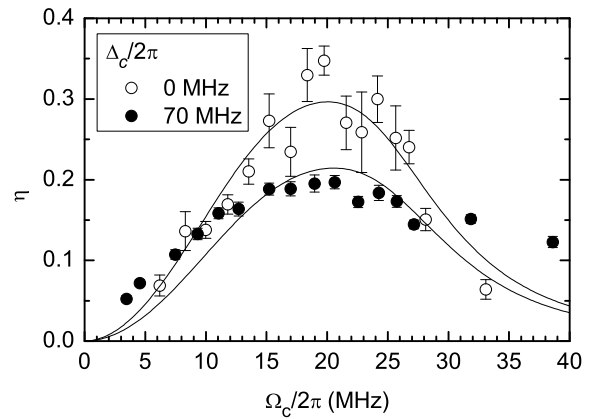}
\caption{
\label{fig-efficiency-vs-power}
Write-read efficiency $\eta$ vs.\ Rabi frequency of the control light $\Omega_c$. All data were taken at the two-photon resonance. Data taken on the single-photon resonance $\Delta_c=0$ ($\circ$) display a clear maximum as a function of $\Omega_c$. Data taken at a single-photon detuning of $\Delta_c=2\pi\times70$ MHz ($\bullet$) display a lower maximum efficiency. The lines are fits of the simple model from appendix A to the data.
}
\end{figure}

Figure \ref{fig-efficiency-vs-power} shows the experimentally observed dependence of $\eta$ on the Rabi frequency of the control light $\Omega_c$. Data were taken for a storage time of 1 $\mu$s with an incoming probe pulse that has a Gaussian intensity profile
\begin{eqnarray}
\label{I-in}
I_\mathrm{in}(t,z) = I_0 \exp\left(-\frac{1}{2\tau_p^2} \left(t -\frac{z}{c} \right)^2 \right)
.\end{eqnarray}
Here, $I_0$ is the peak intensity and $\tau_p$ is the temporal rms width of the intensity. Data in Fig.\ \ref{fig-efficiency-vs-power} were taken for $\tau_p=94$ ns. The control light was turned off $t_0=230$ ns after the maximum probe intensity entered the medium.

Data taken at the single-photon resonance $\Delta_c=0$ ($\circ$) show a maximum of $\eta$ at $\Omega_c\sim 2\pi\times 20$ MHz. This value of $\Omega_c$ agrees fairly well with the prediction of the simple model developed in appendix A. The observed maximum efficiency of $\eta\sim 30$ \%, however, is a factor of $\sim2$ lower than the expectation from the simple model. We attribute this discrepancy to the simplicity of the model and to experimental issues, such as inaccuracies in the determination of the experimental parameters. Note that because of the results of Fig.\ \ref{fig-efficiency-vs-power}, the data in Figs.\ \ref{fig-Faraday} and \ref{fig-alpha} were taken at $\Omega_c=2\pi\times 20$ MHz, where $\eta$ is maximized.

We find experimentally that the write-read efficiency $\eta$ is increased if we slowly decrease the intensity of the control beam while the probe pulse enters the medium. This observation agrees with a more rigorous optimization of $\eta$ for a homogeneous medium in Ref.\ \cite{novikova:07}. Hence, all data in Fig.\ \ref{fig-efficiency-vs-power} were taken with a linear ramp of the intensity of the control beam applied, with a ramp speed that is experimentally found to maximize $\eta$. This ramp is not included in our simple model. For storage, the control light is on for 300 ns because prior to this, there is no probe light inside the medium. The horizontal axis in the figure shows the Rabi frequency corresponding to the time-averaged value of the control intensity during this 300 ns control pulse.

In a measurement independent from Fig.\ \ref{fig-efficiency-vs-power}, we achieved a write-read efficiency of $\eta=(53\pm5)$ \%. The gain in efficiency compared to the data in Fig.\ \ref{fig-efficiency-vs-power} resulted from two changes in the experimental procedure. First, we removed the filter cavity and the mode-filtering fiber after the BEC and, second, we truncated the Gaussian input probe pulse in time, such that it misses exactly that part of its falling edge that cannot be stored anyway because it reaches the BEC after $\Omega_c$ is already ramped to zero.

For reasons discussed in the introduction, our experiments in Ref.\ \cite{lettner:11} had to be operated at a single-photon detuning of $\Delta_c=2\pi\times70$ MHz. An investigation of $\eta$ at this detuning was therefore necessary. Experimental results ($\bullet$) are shown in Fig.\ \ref{fig-efficiency-vs-power}. These data display a maximum value of $\eta \sim 20$ \% at $\Omega_c\sim 2\pi\times 20$ MHz. The value of $\Omega_c$ at which the maximum occurs is essentially identical to the data at the single-photon resonance, whereas the maximum efficiency is further reduced. The physical origin of this reduction is discussed in appendix \ref{sub-app-detuning}.

The lines in Fig.\ \ref{fig-efficiency-vs-power} show fits to the data, based on the simple model developed in appendix A. More precisely, the dash-dotted line from Fig.\ \ref{fig-efficiency-theory} was taken and two fit parameters were introduced, each representing a linear scaling, one for $\eta$ and one for $\Omega_c$.

\subsection{Dependence on the storage time}
\label{sub-eta-time}

Our experiments in Ref.\ \cite{lettner:11} also required an investigation of the time scale on which $\eta$ decays during storage. Thermal motion is known to be the limiting physical effect in many experiments. Using a BEC or an optical lattice, however, thermal motion can be suppressed drastically, resulting in a very slow decay of $\eta(t_\mathrm{store})$ \cite{schnorrberger:09, zhang:09, Dudin:10}. Unlike those experiments, our experiment does not use co-propagating probe and control beams. Instead, the level scheme shown in Fig.\ \ref{fig-levels} requires the two beams to propagate perpendicularly to each other. The resulting differential photon recoil is much larger than for co-propagating beams. In our experiment, the lifetime of $\eta(t_\mathrm{store})$ is predominantly limited by this recoil, similar to Ref.\ \cite{ginsberg:07}.

As a result of the photon recoil, atoms in hyperfine states $F=1$ and $F=2$ move relative to each other. If after $t_\mathrm{store}$ these two atomic clouds do not overlap any more, the retrieval does not produce a directed beam and hardly any signal reaches the detector. In our experiment, the control and probe beams propagate along the $y$ and $z$ axes, respectively. Hence, the differential photon recoil incurred in the Raman transition is directed in the $yz$ plane, where the BEC is symmetric with Thomas-Fermi radii $R_y=R_z=25$ $\mu$m.

The single-mode fiber between the BEC and the detector poses an additional constraint, also related to the photon recoil. Not only do the two atomic clouds need to overlap, the emitted light must also match the transverse mode of the single-mode fiber, resulting in a spatial filtering in the $xy$ plane. The fiber is pretty well mode matched to the incoming probe beam, which has a beam waist of $w=8$ $\mu$m, thus setting a length scale for the spatial filtering that is more stringent along $y$ than the Thomas-Fermi radius $R_y$.

To obtain a simple estimate for $\eta(t_\mathrm{store})$, we approximate the medium as homogeneous, which is justified by $w\ll R_y$ and $w\ll R_z$. In this approximation, only the recoil along the $y$ axis is relevant, i.e.\ the photon recoil of the probe laser is irrelevant. As the mode of the optical fiber has a Gaussian transverse profile, the decay of $\eta(t_\mathrm{store})$ is expected to be Gaussian
\begin{eqnarray}
\label{eta-Gauss}
\eta(t_\mathrm{store})
= \eta(0) e^{-t_\mathrm{store}^2/2\sigma_\eta^2}
\end{eqnarray}
with a $e^{-1/2}$ time $\sigma_\eta=mw/\sqrt2 \hbar k_c=1.0$ ms, where $m$ is the atomic mass and $k_c=\omega_c/c$ is the wave vector of the control beam. Note that the factor $\sqrt2$ here has nothing to do with the modulus of the differential photon recoil.

\begin{figure}[t!]
\includegraphics[width=0.9\columnwidth]{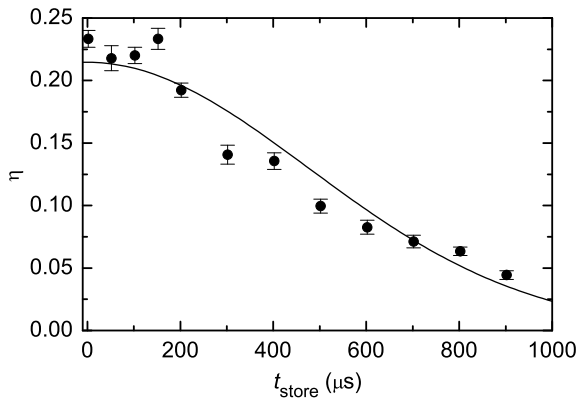}
\caption{
\label{fig-EtaVsTimePureBEC}
Decay of the write-read efficiency $\eta$ as a function of storage time in a pure BEC. The line shows a Gaussian fit according to Eq.\ \eqref{eta-Gauss}, yielding a best-fit value $\sigma_\eta=0.48$ ms for the $e^{-1/2}$ time of $\eta$. The dominant mechanism that sets this time scale is given by spatial filtering caused by the single-mode fiber combined with photon recoil incurred during the storage process.
}
\end{figure}

Figure \ref{fig-EtaVsTimePureBEC} shows experimental data for $\eta(t_\mathrm{store})$, recorded for an essentially pure BEC and a single-photon detuning of $\Delta_c=2\pi\times 70$ MHz. The line shows a fit of Eq.\ \eqref{eta-Gauss} to the data, yielding a best-fit value of $\sigma_\eta=0.48$ ms. This is a factor of $\sim 2$ less than expected, probably owing to the simplicity of the model. Note that the decay of $\eta$ observed here is unproblematic for our experiment in Ref.\ \cite{lettner:11} where we took data only for $t_\mathrm{store}\leq300$ $\mu$s. Furthermore, the decay of $\alpha$ observed in Fig.\ \ref{fig-alpha} is much slower than the decay of $\eta$ observed in Fig.\ \ref{fig-EtaVsTimePureBEC}. This means that at long storage times, only very little light is retrieved but it still has the correct polarization.

Achieving a much slower decay of $\eta$ during storage would be possible when using co-propagating or almost co-propagating beams, as mentioned above. This would be incompatible with the present atomic level scheme. But a conversion of the polarization qubit into different wave vectors of the control light, as in Ref.\ \cite{Zhang:11}, could solve this problem. However, our experiments in Ref.\ \cite{lettner:11} would not immediately profit from a slower decay of $\eta$ for two reasons. First, very long storage times would drastically slow down the rate at which write-read cycles can be repeated, which would result in an unrealistically long data acquisition time for the complete experiment, due to low count rates. Second, in the setup used in Ref.\ \cite{lettner:11}, the limiting factor when extending the storage time was the deterioration of the fidelity due to magnetic field noise acting on the single atom inside the high-finesse cavity.

\begin{figure}[t!]
\includegraphics[width=0.9\columnwidth]{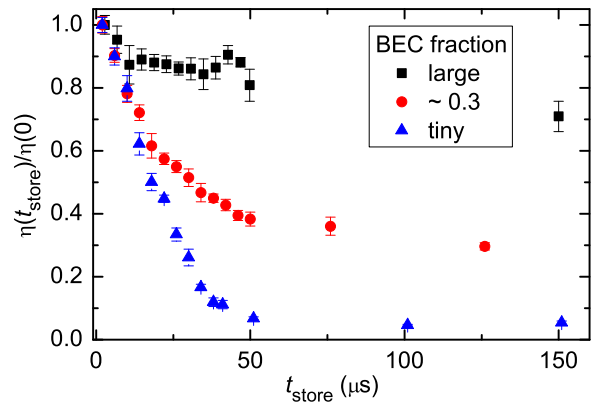}
\caption{
\label{fig-EtaVsTimeBimodal}
(Color online) Decay of the normalized write-read efficiency $\eta$ in the presence of a noticeable uncondensed fraction of the gas. Thermal motion causes a rapid initial decay of $\eta$. The decay settles to the long-lived level of $\eta$ that is caused by the BEC fraction. The data sets were taken for different BEC fractions and, correspondingly, for different temperatures.
}
\end{figure}

To illustrate how our experiments profit from the use of a BEC, we deliberately prepare an atomic gas with a noticeable uncondensed fraction. Figure \ref{fig-EtaVsTimeBimodal} shows that in this case $\eta(t_\mathrm{store})$ decays much faster. More specifically, $\eta$ is the sum of two contributions, one from the BEC and one from the uncondensed fraction. These two contributions to $\eta$ decay on quite different time scales. On the time scale shown in Fig.\ \ref{fig-EtaVsTimeBimodal} the contribution of the uncondensed fraction decays almost completely, whereas the contribution of the BEC is essentially constant. The overall decay of $\eta$ is sensitive to the first-order spatial coherence function of the gas \cite{naraschewski:99, ginsberg:07}. A bimodal decay of the first-order coherence similar to Fig.\ \ref{fig-EtaVsTimeBimodal} was previously observed in Ref.\ \cite{bloch:00} using a different technique.

Comparing the data in Fig.\ \ref{fig-EtaVsTimeBimodal} to the temperatures extracted from the size of the uncondensed fraction in time-of-flight images, we confirm that $\lambda_\mathrm{dB}/v_\mathrm{rel}$ can be used as a coarse estimate for the time scale of the decay of $\eta$ caused by the uncondensed fraction of the gas. Here, $\bm v_\mathrm{rel}=\hbar (\bm k_p-\bm k_c)/m$ is the velocity of the $F=2$ atoms relative to the $F=1$ atoms during $t_\mathrm{store}$, $\bm k_p$ is the probe-light wave-vector, $\lambda_{dB}=\sqrt{2\pi\hbar^2/mk_BT}$ is the thermal de-Broglie wavelength, $T$ is the temperature, and $k_B$ is the Boltzmann constant. The data in Fig.\ \ref{fig-EtaVsTimeBimodal} were taken with a waist of the probe beam of $w=30$ $\mu$m and after removing the filter cavity as well as the single-mode fiber between the BEC and the detector. Without these modifications, $\eta$ would be sensitive only to the central region of the gas, where the uncondensed fraction contributes less.

\section{Conclusion}

To conclude, we characterized and optimized the BEC as a quantum memory and showed that a write-read efficiency above 50 \% can be reached. Its decay over storage time results from the differential photon recoil in the Raman transfer combined with spatial filtering of the retrieved light. This could be mitigated using co-propagating light beams. We also showed that the mapping between photonic and atomic qubit has an impressive average process fidelity. Its decay over storage time is due to magnetic-field noise and is suppressed by appropriate techniques.

\acknowledgments

We thank E.\ Figueroa and S.\ Ritter for discussions. This work was supported by the German Excellence Initiative through the Nanosystems Initiative Munich and by the Deutsche Forschungsgemeinschaft through SFB 631.

\appendix

\section{Modeling the efficiency}
\label{app}

In this appendix, we derive a simple estimate for the write-read efficiency $\eta$. This estimate was used for fitting to the experimental data in Fig.\ \ref{fig-efficiency-vs-power}. To set the stage for this calculation, Sec.\ \ref{sub-app-theory} briefly summarizes the theoretical background from the literature. We then develop a simple, largely analytic model for $\eta$ in Sec.\ \ref{sub-app-simple}. In Sec.\ \ref{sub-app-detuning}, we discuss why a large single-photon detuning reduces $\eta$.

\subsection{Theoretical background}
\label{sub-app-theory}

Our notation largely follows Ref.\ \cite{fleischhauer:05}, except for the sign of all detunings. We denote the probe and control Rabi frequencies as $\Omega_p$ and $\Omega_c$ and the corresponding detunings as $\Delta_p=\omega_p-\omega_{p,\mathrm{res}}$ and $\Delta_c=\omega_c-\omega_{c,\mathrm{res}}$, where $\omega_p$ and $\omega_c$ are the angular frequencies of the light fields and $\omega_{p,\mathrm{res}}$ and $\omega_{c,\mathrm{res}}$ are the corresponding atomic resonances, respectively. We consider the regime of small probe intensity and neglect dephasing. The propagation of the probe light can then be described by the linear susceptibility \cite{fleischhauer:05}
\begin{eqnarray}
\label{chi}
\chi = \chi_0
\frac{2\delta_2 \Gamma}{\Omega_c^2-4\delta_2(\Delta_c+\delta_2)-2i\delta_2\Gamma}
\end{eqnarray}
with
\begin{eqnarray}
\chi_0 = \frac{n_\mathrm{gr}\Omega_c^2}{\omega_p \Gamma}
\end{eqnarray}
and the group index \cite{fleischhauer:05}
\begin{eqnarray}
n_\mathrm{gr} = \frac{\Gamma_p}{\Omega_c^2} \varrho \sigma c
.\end{eqnarray}
Here $\delta_2=\Delta_p-\Delta_c$ is the two-photon detuning, $\Gamma$ is the total decay rate of the excited state, $\Gamma_p$ is the partial decay rate of the excited state into the ground state involved in the probe transition, $c$ is the vacuum speed of light, $\varrho$ is the particle density, and $\sigma=3\lambda_p^2/2\pi$ is the resonant light scattering cross section for a cycling transition at wavelength $\lambda_p=2\pi c/\omega_p$.

If the two-photon detuning $\delta_2$ is small, then one can expand ${\rm Re}(\chi)$ and ${\rm Im}(\chi)$ to lowest non-vanishing order in $\delta_2$, yielding
\begin{subequations}
\label{chi-approx}
\begin{eqnarray}
\label{Re-chi-approx}
\mathrm{Re}(\chi) &=& \chi_0 \frac{2\Gamma}{\Omega_c^2} \delta_2 +\mathcal O(\delta_2^2)
,\\
\label{Im-chi-approx}
\mathrm{Im}(\chi) &=& \chi_0 \left( \frac{2\Gamma}{\Omega_c^2} \delta_2\right)^2 +\mathcal O(\delta_2^3)
.\end{eqnarray}
\end{subequations}
Note that this is independent of $\Delta_c$. The group velocity for probe light can be calculated from
Eq.\ \eqref{Re-chi-approx}, yielding \cite{fleischhauer:05}
\begin{eqnarray}
\label{v-gr}
v_\mathrm{gr} = \frac{c}{1+n_\mathrm{gr}}
.\end{eqnarray}
We consider a medium that extends from $z=0$ to $z=L$ with a density $\varrho(z)$ which varies along the propagation direction $z$ of the probe beam. If $\Omega_c$ is constant in time, the pulse delay after propagation through the complete medium follows immediately from Eq.\ \eqref{v-gr}, yielding \cite{fleischhauer:05}
\begin{eqnarray}
\label{tau-d}
\tau_d(L) = \int_0^L dz \frac{n_\mathrm{gr}(z)}{c}
= \frac{\Gamma}{\Omega_c^2} d_p(L)
,
\end{eqnarray}
where
\begin{eqnarray}
\label{d-p}
d_p (L)= \int_0^L dz \frac{\Gamma_p}{\Gamma} \sigma \varrho(z)
\end{eqnarray}
denotes the optical depth seen by the probe light.

Irreversible absorption of the probe light inside the medium can be a serious issue. This is avoided if all relevant frequency components of the probe light are close to the two-photon resonance, $\delta_2=0$. Under this condition, Eq.\ \eqref{Im-chi-approx} yields a Gaussian EIT intensity transmission window in frequency space with rms width $\sigma_\mathrm{trans}=\Delta\omega_\mathrm{trans}/\sqrt8$ with \cite{fleischhauer:05}
\begin{eqnarray}
\label{Delta-omega-trans}
\Delta\omega_\mathrm{trans}(L)
= \frac{\Omega_c^2}{\Gamma \sqrt{d_p(L)}}
.\end{eqnarray}
If $\Delta \omega_p$ denotes the typical width of the frequency spectrum of the probe pulse, then the condition for small absorption reads $\Delta \omega_p\ll \Delta\omega_\mathrm{trans}(L)$. Due to the Fourier limit, the typical duration $\tau_p$ of the pulse is related to $\Delta\omega_p$ by $\tau_p \Delta\omega_p \sim 1$. Combining this with Eqs.\ \eqref{tau-d} and \eqref{Delta-omega-trans}, the condition for small absorption can be rewritten as
\begin{eqnarray}
\label{tau-tau}
\frac{\tau_d(L)}{\tau_p} \ll \sqrt{d_p(L)}
.\end{eqnarray}
Obviously, fully compressing the pulse longitudinally into the medium requires $\tau_d(L)/\tau_p>1$. If one simultaneously wants to avoid absorption, then according to Eq.\ \eqref{tau-tau} one needs $d_p(L)\gg 1$. The requirement of large optical depth is independent of $\Omega_c$. But in the experiment, $\Omega_c$ must be adapted to the values of $\tau_p$ and $d_p(L)$, as discussed now.

\subsection{Simple estimate for the efficiency}
\label{sub-app-simple}

While various numerical models for a thorough analysis of $\eta$ have been published, we find it useful to complement these elaborate models with a much simpler model that captures only part of the physics but gives a quick estimate for the efficiency. Our model assumes that the probe pulse is Gaussian in the time domain. Using Eq.\ \eqref{chi-approx}, the intensity of the propagating probe pulse can be approximated as
\begin{eqnarray}
\label{I-t-z}
I(t,z) = I_0(z) \exp\left(-\frac{1}{2\tau_p^2(z)} \left(t - \tau_d(z) -\frac{z}{c} \right)^2 \right)
,\quad
\end{eqnarray}
where $I_0$ is the peak intensity, $\tau_p$ is the temporal rms width of the intensity, and $\tau_d$ is given in Eq.\ \eqref{tau-d}. Note that Eq.\ \eqref{I-t-z} is valid for all $z$. It simplifies to Eq.\ \eqref{I-in} for those values of $z$ where the probe light has not yet entered the medium.

Our simple model separately addresses the issues of insufficient pulse compression and irreversible absorption. First, we ignore irreversible absorption. This makes $I_0$ and $\tau_p$ independent of $z$. We assume that $\Omega_c$ is switched off at a time $t_0$ and assume that the fraction of the light that is inside the medium at this moment is stored and subsequently retrieved. This yields a write-read efficiency
\begin{eqnarray}
\eta_\mathrm{comp} = \frac12 \left[ \mathrm{erf} \left( \frac{t_0}{\sqrt2 \tau_p} \right)
- \mathrm{erf} \left( \frac{t_0-\tau_d(L)-L/c}{\sqrt2 \tau_p} \right) \right]
\nonumber
\\
\label{eta-comp}
\end{eqnarray}
with the error function $\mathrm{erf}(x) =(2/\sqrt\pi) \int_0^x du \exp(-u^2)$. This result for the efficiency quantifies how well the pulse is compressed into the medium. Note that $n_\mathrm{gr}\gg1$ implies $\tau_d(L)\gg L/c$.

Second, we turn to irreversible absorption. Here, we consider a situation in which $\Omega_c$ is constant in time, implying that no storage takes place. The fraction $\eta_\mathrm{trans}$ of the pulse energy that is transmitted through the medium is calculated easily in the frequency domain. Based on Eqs.\ \eqref{Im-chi-approx} and \eqref{I-t-z}, we obtain
\begin{eqnarray}
\label{eta-trans}
\eta_\mathrm{trans}
= \left( 1+ \frac{2}{(\tau_p \Delta\omega_\mathrm{trans}(L))^2} \right)^{-1/2}
\end{eqnarray}
with $\Delta\omega_\mathrm{trans}$ from Eq.\ \eqref{Delta-omega-trans}. This result for the efficiency expresses the issue of irreversible absorption.

\begin{figure}[t!]
\includegraphics[width=0.9\columnwidth]{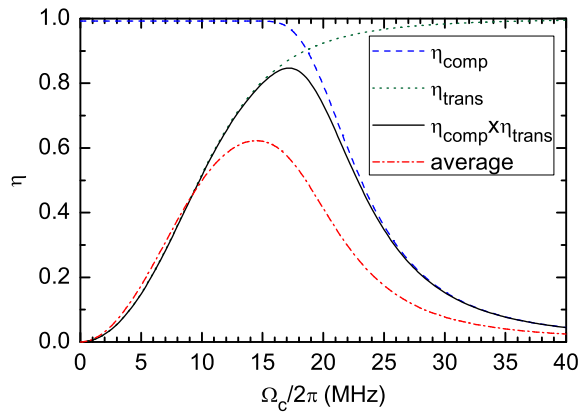}
\caption{
\label{fig-efficiency-theory}
(Color online) Simple theoretical estimate for the write-read efficiency $\eta$ vs.\ Rabi frequency of the control light $\Omega_c$. The dashed, dotted, and solid lines show $\eta_\mathrm{comp}$, $\eta_\mathrm{trans}$, and $\eta_\mathrm{comp}\times \eta_\mathrm{trans}$, respectively. Parameters are $\tau_p=94$ ns, $t_0=230$ ns, $1/\Gamma=26$ ns, and $d_p(L)=127$. The pronounced maximum in the solid line arises because for small $\Omega_c$, the EIT transmission window in frequency space is too narrow for the incoming probe pulse, whereas for large $\Omega_c$, the group velocity is not sufficiently reduced to spatially compress the complete pulse into the BEC. The dash-dotted line shows the result of averaging over the transverse profile of the BEC.
}
\end{figure}

To obtain a simple estimate for the overall efficiency, which must take both effects into account, we simply multiply the two efficiencies from Eqs.\ \eqref{eta-comp} and \eqref{eta-trans}. For the atomic probe transition used in our experiment, we have $\Gamma_p=\Gamma/12$. Combination with the atom number and trap frequencies quoted in Sec.\ \ref{sub-impl-BEC} yields a peak value of $d_p(L)=127$ at $x=y=0$. Results for this optical depth and typical parameters of our experimental pulses are shown in Fig.\ \ref{fig-efficiency-theory}.

In the model, we explored $\eta$ as a function of the two-dimensional parameter space spanned by $\Omega_c$ and $t_0$. Fig.\ \ref{fig-efficiency-theory} shows the dependence on $\Omega_c$ only for that value of $t_0$, for which the global maximum of $\eta$ is reached. A modification of the pulse duration $\tau_p$ would require a re-optimization of $\Omega_c$ and $t_0$. Considering Eqs.\ \eqref{tau-d} and \eqref{Delta-omega-trans}, one finds that Eqs.\ \eqref{eta-comp} and \eqref{eta-trans} remain unchanged if the scalings $\Omega_c\propto1/\sqrt{\tau_d}$ and $t_0\propto\tau_p$ are used. As a consequence, the maximum value of $\eta(\Omega_c,t_0)$ is insensitive to a change in $\tau_d$.

The transverse inhomogeneity of the BEC can be accounted for by calculating a weighted average of $\eta$, with the transverse profile of the probe light intensity as a weighting function
\begin{eqnarray}
\int dx dy \frac{2}{\pi w^2} e^{-2(x^2+y^2)/w^2} \eta(d_p(x,y))
. \end{eqnarray}
We assume a Thomas-Fermi parabola for $\varrho(x,y,z)$ with Thomas-Fermi radii $R_x$, $R_y$, and $R_z$. Calculation of $d_p(x,y)$ by analytic integration over $z$ is straightforward. After a transformation to new coordinates $(\tilde\rho,\varphi)$ with $x=R_x\tilde\rho\cos\varphi$ and $y=R_y\tilde\rho\sin\varphi$, the integral over $\varphi$ can also be solved analytically. The remaining integral over $\tilde\rho$ is easily computed numerically. The result is shown as a dash-dotted line in Fig.\ \ref{fig-efficiency-theory}. This line predicts a maximum of $\eta\sim60$ \% at $\Omega_c=2\pi\times15$ MHz. For this value of $\Omega_c$ and for $x=y=0$, the theory yields estimated values of $\chi_0=0.5$, $n_\mathrm{gr} = 5\times 10^6$, $\tau_d(L) = 550$ ns, and $\Delta\omega_\mathrm{trans}(L) = 2\pi\times 3.3$ MHz.

\begin{figure}[t!]
\includegraphics[width=0.9\columnwidth]{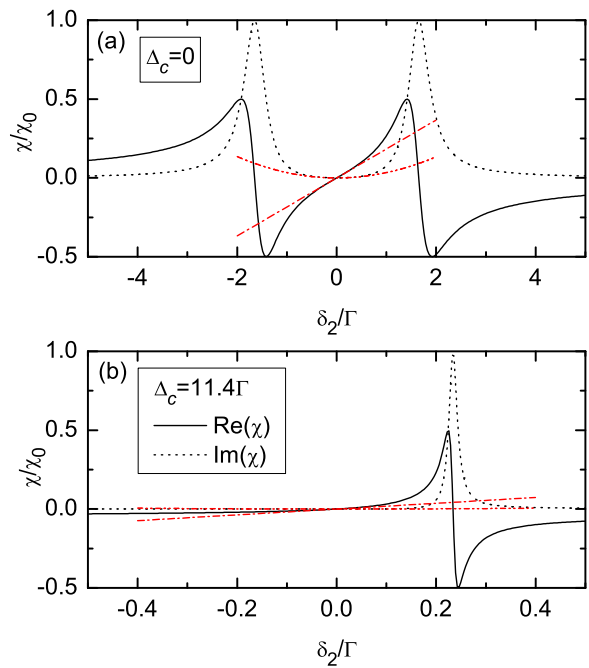}
\caption{
\label{fig-chi}
(Color online) Linear susceptibility $\chi$ as a function of the two-photon detuning $\delta_2$. The solid and dotted lines show the predictions of Eq.\ \eqref{chi} for the real and imaginary parts of $\chi$ at $\Omega_c = 3.3 \Gamma = 2\pi\times 20$ MHz. (a) At the single photon resonance, $\Delta_c=0$. (b) At a single photon detuning of $\Delta_c=11.4 \Gamma = 2\pi\times 70$ MHz. Dash-dotted lines (red) show the lowest-order approximations of Eq.\ \eqref{chi-approx}. The range where these approximations become poor sets an upper bound for the frequency range which is useful for light storage. This range is much narrower in (b) than in (a). This explains the reduction of the efficiency observed at $\Delta_c=2\pi\times 70$ MHz. Note that the scales on the horizontal axes differ by one order of magnitude.
}
\end{figure}

The absorption represents a filter in frequency space. Due to the Fourier limit, this causes an increase of the temporal pulse width $\tau_p$ for increasing $z$, thus complicating a more rigorous calculation of $\eta_\mathrm{comp}$. We can overestimate the effect of this increase of $\tau_p(z)$ when using the constant value $\tau_p(L)$ instead of $\tau_p(0)$ in calculating $\eta_\mathrm{comp}$. We find that for the parameters of Fig.\ \ref{fig-efficiency-theory}, this has little effect. Our model also neglects that $\tau_p(z)$ should increase due to dispersion caused by $\mathrm{Re}(\chi)$. This increase is given by $\tau_p^2(z)=\tau_p^2(0)+[\tau_d(z)/\tau_p(0)\omega_p]^2$, which is negligible for the parameters of our experiment.

The simple model developed here neglects that the spatial cutting during storage broadens the frequency spectrum of the pulse, thus increasing the absorption after retrieval. The model also neglects that the transverse inhomogeneity of the medium might cause a deformation of the wavefronts, resulting in effects such as focussing of the probe beam. In addition, this model is fully based on Eq.\ \eqref{chi-approx}, instead of Eq.\ \eqref{chi}. If the probe pulse is so broad in frequency space that Eq.\ \eqref{chi-approx} is not a good approximation, then the applicability of the model is questionable. Nonetheless, the prediction of the model agrees reasonably well with our experimental data in Fig.~\ref{fig-efficiency-vs-power}.

\subsection{Reduced efficiency at large single-photon detuning}
\label{sub-app-detuning}

To understand the physical origin of the reduction of $\eta$ at large single-photon detuning, we investigate the frequency range that is useful for storing light. An upper limit for this frequency range is set by the frequency range within which Eq.\ \eqref{chi-approx} is a good approximation. Using Eq.\ \eqref{chi}, one can easily show that the maxima of $\mathrm{Im}(\chi)$ are located at
\begin{eqnarray}
\delta_2 = \frac12\left(-\Delta_c \pm \sqrt{\Delta_c^2 + \Omega_c^2} \right)
.\end{eqnarray}
The maximum nearest to $\delta_2=0$ clearly sets an overoptimistic upper bound for the useful frequency range for light storage. For $|\Delta_c|\gg \Omega_c$, the nearest maximum lies at $\delta_2 \approx \Omega_c^2/4 \Delta_c$. Comparison with Eq.\ \eqref{Delta-omega-trans} shows that the useful frequency range is much narrower than $\Delta\omega_\mathrm{trans}$ unless $4|\Delta_c|/\Gamma \sqrt{d_p}\ll 1$. Our experiment is operated at $4\Delta_c/\Gamma \sqrt{d_p} \approx 4$ so that this issue is obviously relevant. This reduction of the useful frequency range is to be contrasted with the delay $\tau_d$ in Eq.\ \eqref{tau-d} which is independent of $\Delta_c$. As a result, the overall efficiency is reduced.

To further illustrate this point, we show the dependence of $\chi$ on $\delta_2$ in Fig.\ \ref{fig-chi}. Parts (a) and (b) correspond to $\Delta_c=0$ and $\Delta_c=2\pi\times 70$ MHz, respectively. Note the different scales on the horizontal axes. This figure clearly illustrates that the frequency range over which Eq.\ \eqref{chi-approx} is a good approximation differs drastically between the two cases.

The light pulses that we store at $\Delta_c=2\pi\times 70$ MHz have the same spectral width as for $\Delta_c=0$. For $\Delta_c=2\pi\times 70$ MHz a considerable part of the frequency components of the light therefore samples the frequency range where Eq.\ \eqref{chi-approx} is not a good approximation. Fig.\ \ref{fig-chi}(b) shows that for components with negative $\delta_2$, the value of $d\mathrm{Re}(\chi)/d\delta_2$ is reduced, resulting in a faster group velocity, which is disadvantageous. On the other hand, for positive $\delta_2$ absorption can be substantial and $d\mathrm{Re}(\chi)/d\delta_2$ can even change sign, thus not creating slow light. These problems qualitatively explain the reduced write-read efficiency that we observe experimentally for $\Delta_c=2\pi\times 70$ MHz.

\end{document}